%% file: ForeSeer_CIKM2023.tex
\documentclass[sigconf]{acmart}
\usepackage{latexsym}
\usepackage{multirow}
\usepackage{amsmath}
\usepackage{graphicx}[pdflatex]

\AtBeginDocument{%
  \providecommand\BibTeX{{%
    \normalfont B\kern-0.5em{\scshape i\kern-0.25em b}\kern-0.8em\TeX}}}

\makeatletter
\gdef\@copyrightpermission{
  \begin{minipage}{0.3\columnwidth}
   \href{https://creativecommons.org/licenses/by/4.0/}{\includegraphics[width=0.90\textwidth]{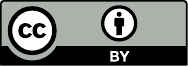}}
  \end{minipage}\hfill
  \begin{minipage}{0.7\columnwidth}
   \href{https://creativecommons.org/licenses/by/4.0/}{This work is licensed under a Creative Commons Attribution International 4.0 License.}
  \end{minipage}
  \vspace{5pt}
}
\makeatother

\copyrightyear{2023}
\acmYear{2023}
\setcopyright{rightsretained}
\acmConference[CIKM '23] {Proceedings of the 32nd ACM International Conference on Information and Knowledge Management}{October 21--25, 2023}{Birmingham, United Kingdom.}
\acmBooktitle{Proceedings of the 32nd ACM International Conference on Information and Knowledge Management (CIKM '23), October 21--25, 2023, Birmingham, United Kingdom}
\acmPrice{}
\acmISBN{979-8-4007-0124-5/23/10}
\acmDOI{10.1145/3583780.3614887}	

\begin{document}

\title{ForeSeer: Product Aspect Forecasting Using Temporal Graph Embedding}



\author{Zixuan Liu}
\email{zucksliu@cs.washington.edu}
\authornote{This research was conducted during the author's internship at Amazon.}
\orcid{0000-0002-5730-9987}
\affiliation{%
  \institution{University of Washington}
  \streetaddress{P.O. Box 1212}
  \city{Seattle}
  \state{WA}
  \country{USA}
  \postcode{98105}
}

\author{Gaurush Hiranadani}
\email{hgaurush@amazon.com}
\affiliation{%
  \institution{Amazon}
  \city{Palo Alto}
  \state{CA}
  \country{USA}}
  \orcid{0009-0002-5718-8959}

\author{Kun Qian}
\email{qianku@amazon.com}
\affiliation{%
  \institution{Amazon}
  \city{Palo Alto}
  \state{CA}
  \country{USA}}
  \orcid{0000-0002-9063-102X}

\author{Edward W. Huang}
\email{ewhuang@amazon.com}
\affiliation{%
  \institution{Amazon}
  \city{Palo Alto}
  \state{CA}
  \country{USA}}
  \orcid{0000-0002-4461-8545}

\author{Yi Xu}
\email{yxaamzn@amazon.com}
\affiliation{%
  \institution{Amazon}
  \city{Seattle}
  \state{WA}
  \country{USA}}
  \orcid{0000-0002-0604-8481}

\author{Belinda Zeng}
\email{zengb@amazon.com}
\affiliation{%
  \institution{Amazon}
  \city{Seattle}
  \state{WA}
  \country{USA}}
  \orcid{0009-0002-5446-0523}

\author{Karthik Subbian}
\email{ksubbian@amazon.com}
\affiliation{%
  \institution{Amazon}
  \city{Palo Alto}
  \state{CA}
  \country{USA}}
  \orcid{0000-0002-9023-2248}

\author{Sheng Wang}
\email{swang@cs.washington.edu}
\affiliation{%
  \institution{University of Washington}
  \city{Seattle}
  \state{WA}
  \country{USA}}
  \orcid{0000-0002-0439-5199}

\renewcommand{\shortauthors}{Zixuan Liu et al.}

\ccsdesc[500]{Information systems~Data mining}
\ccsdesc[500]{Information systems~Information retrieval}
\ccsdesc[300]{Information systems~Language models}
\ccsdesc[500]{Computing methodologies~Machine learning}
\ccsdesc[500]{Computing methodologies~Information extraction}
\ccsdesc[500]{Computing methodologies~Semantic networks}
\ccsdesc[500]{Computing methodologies~Temporal reasoning}
\ccsdesc[300]{Computing methodologies~Transfer learning}
\keywords{Aspect forecasting, textual mining, temporal graph embedding, multi-time forecasting, information extraction, contrastive learning}

\input{main_paper/abstract}
\maketitle

\input{main_paper/introduction}
\input{main_paper/problem_setting}

\input{main_paper/methods}
\input{main_paper/experiments}
\input{main_paper/related_works}

\input{main_paper/conclusion}




\bibliographystyle{ACM-Reference-Format}
\bibliography{sample-base}
\end{document}

%% file: main_paper/abstract.tex
\begin{abstract}
Developing text mining approaches to mine aspects from customer reviews has been well-studied due to its importance in understanding customer needs and product attributes. In contrast, it remains unclear how to predict the future emerging aspects of a new product that currently has little review information. This task, which we named product aspect forecasting, is critical for recommending new products, but also challenging because of the missing reviews. Here, we propose ForeSeer, a novel textual mining and product embedding approach progressively trained on temporal product graphs for this novel product aspect forecasting task. ForeSeer transfers reviews from similar products on a large product graph and exploits these reviews to predict aspects that might emerge in future reviews. A key novelty of our method is to jointly provide review, product, and aspect embeddings that are both time-sensitive and less affected by extremely imbalanced aspect frequencies. 
We evaluated ForeSeer on a real-world product review system containing 11,536,382 reviews and 11,000 products over 3 years. We observe that ForeSeer substantially outperformed existing approaches with at least 49.1\% AUPRC improvement under the real setting where aspect associations are not given. ForeSeer further improves future link prediction on the product graph and the review aspect association prediction. Collectively, Foreseer offers a novel framework for review forecasting by effectively integrating review text, product network, and temporal information, opening up new avenues for online shopping recommendation and e-commerce applications. 
\end{abstract}

%% file: main_paper/introduction.tex
\section{Introduction}
Customer reviews reflect the properties of the product and the needs of the customer on online shopping systems \cite{kim2018analyzing, zhao2019predicting}. As a result, one widely studied task is to extract descriptive keywords from customer reviews, such as "not greasy" and "soft inside". These descriptive keywords, which are referred to as aspects in the literature, are later used as key features for various applications that are beneficial to buyers making decisions and sellers improving the market, such as customer behavior prediction \cite{rana2022classifying}, sentiment analysis \cite{mittal2022determining}, opinion mining \cite{xu2011mining, kumar2016opinion}, product rating \cite{liu2019assessing}, and marketing refinement \cite{verma2021artificial}. To tackle this problem, many text mining methods have been proposed to mine such aspects from customer reviews \cite{xu2013aspect, yi2022informational, sharma2014mining, dadhich2021social, kpiebaareh2022generic, jeyapriya2015extracting, chakraborty2022aspect, zhang2021discovering, zhang2022oa}. 

\begin{figure}[!h]
\centering
 \includegraphics[width=1\linewidth]{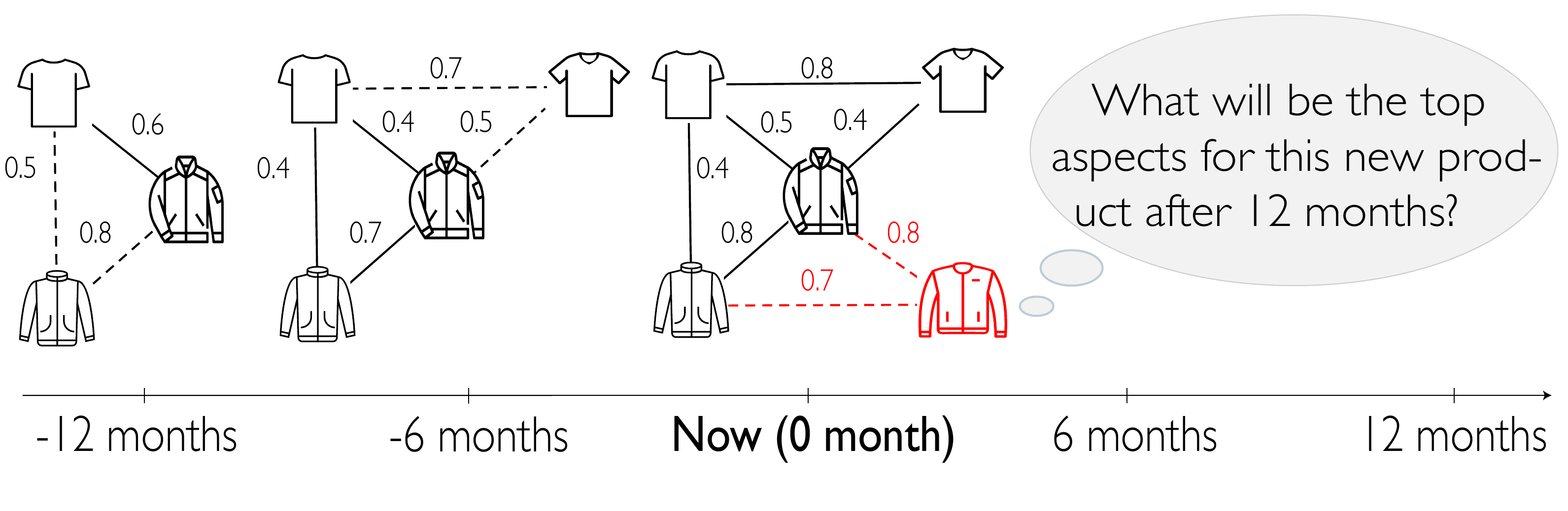}
\caption{Problem setting of aspect forecasting. The product graph is evolving because new product nodes and edges are added to the graph. The edge weights also change over time. }
\label{fig:split_figure1}

\end{figure}

Despite its usefulness, however, customer reviews are also subjective and sometimes biased, therefore mining effective aspects from reviews for a product requires adequate numbers of reviews. For a new product with limited customer reviews, aspect mining methods will suffer from starting cold and mining inaccurate and noisy aspects. In this paper, we aim at a novel \textit{product aspect forecasting} task, by forecasting the top aspects in the future for products with inadequate reviews (Fig. \ref{fig:split_figure1}), i.e., \textit{what aspects will the customer use to describe a new product after six months or three years?} To the best of our knowledge, product aspect forecasting has not been studied in the literature and is one step further than traditional aspect extraction tasks. The most related task is zero-shot aspect-based sentiment analysis \cite{ jebbara2019zero,shu2022zero, deng2022zero}, but their frameworks are restricted to sentiment analysis or still require multilingual supervision. The recent progress in large language models such as ChatGPT \cite{ouyang2022training} or GPT-4 \cite{bubeck2023sparks} may also open a door for improved zero-shot aspect extraction performance, but besides the heavy computational costs using them, the limited availability of reviews of new products still restricts the model from forecasting their top aspects in the future.

\setlength{\abovecaptionskip}{-0.05cm}
\setlength{\belowcaptionskip}{-0.1cm}
\vspace{-4mm}

\vspace{-2mm}

\begin{figure}[!ht]
\vspace{-1mm}
\centering
 \includegraphics[width=\linewidth]{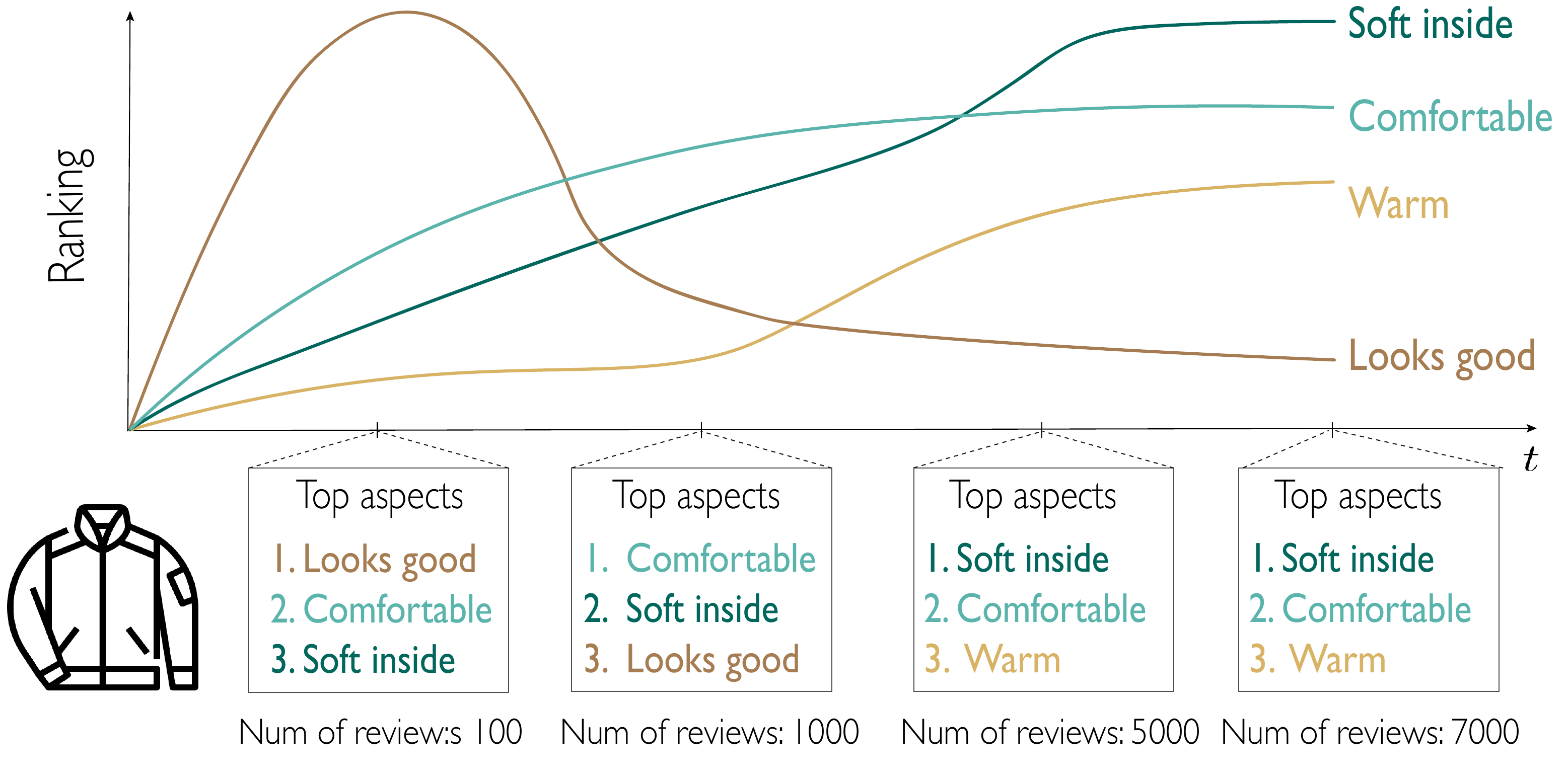}
\caption{The key idea of ForeSeer is to find and embed similar products dynamically and propagate their top aspects mined from reviews to the new item to forecast future aspects. }
\label{fig:split_figure1b}
\end{figure}

In this paper, we formulate the product aspect forecasting task as a multi-future top-K aspect retrieving problem. An intuitive solution to the problem is to find similar products with longer histories and infer based on their received feedback, just like a customer would search for the reviews of similar older products that have the same functionality, brand, or same style before making the decision. Motivated by this, in this paper, we propose a three-stage framework ForeSeer to address it. ForeSeer is a flexible textual mining and product embedding framework progressively trained on temporal product graphs. The key idea is to find similar older products from the temporal product graph and propagate potential aspects mined from their reviews to the new item. This enables forecasting currently under-estimated aspects which might be frequently mentioned in the future (Fig. \ref{fig:split_figure1b}). ForeSeer has three core steps, contrastive review-aspect association learning, progressive temporal graph-based product embedding, and aspect-guided product embedding refinement with temporal information. The first step efficiently captures the static semantic relationship between review texts and aspects, and the second step together captures the evolving graphical product similarity information and how products aggregate reviews over time. In the third step, we design a novel product-aspect temporal attention (PATA) module to help adjust the product embeddings guided by aspect embeddings for multi-length future times forecasting.  


We evaluated ForeSeer on a real-world customer review dataset, which contains 11,000 products, 11,536,382 reviews, and a product similarity graph constructed using user queries and clicks. Our dataset contains the timestamp for each review and product over three years. We observed substantial improvement against comparison approaches on product aspect forecasting in both settings with accessible estimated or ground truth association. We found the product embeddings derived from ForeSeer show more visible patterns when the timestamp is increasing, suggesting an accurate incorporation of temporal information. In addition to aspect forecasting, ForeSeer achieved prominent performance on multi-time future link prediction and review-aspect association prediction, demonstrating its wide applicability in modeling temporal review information. ForeSeer is developed as, to our knowledge, the first approach to product aspect forecasting and can be broadly applied to other temporal graph mining tasks.

%% file: main_paper/problem_setting.tex
\vspace{-5mm}
\section{Product aspect forecasting}

\subsection{Problem Definition: A new task}
\label{sec:prob_def}

In principle, products receive and aggregate reviews and aspects are mined from the review text pieces, enabling the products to extract and aggregate aspects associations. 
We then understand relationships among the three core components, products, aspects, and reviews in detail, and formally define the product aspect mining and forecasting problem as a top-$K_l$ aspect retrieval problem.

\subsubsection{Product aspect mining from review texts} 
A descriptive aspect $A_l$, is a descriptive textual span extracted from a review text sequence $r_j$ that summarizes a particular attribute or feature of $P_i$, the product $r_j$ belongs to (examples and illustration in Table~\ref{tab:example_ra_asso}). 
Given an aspect list $L=\{A_1, A_2, \dots, A_{|L|}\}$ and $r_j$, 
we define the review-aspect association vector $\mathbf{v}_{j} \in \{0,1\}^{|L|}$, where $v^{l}_{j} = 1$ if $r_{j}$ and $A_l$ are associated. The aspect list can be obtained by supervised or unsupervised mining from reviews \cite{zhang2022oa,yang2022mave}. The longer the list, the more descriptive and diverse aspects can be included in the list. 

A product network $\mathcal{G} = \{\mathcal{P}, \mathcal{E}\}$ is given with a set of product nodes $\mathcal{P}$ and an edge set $\mathcal{E}$ representing the similarities between products. $\mathbf{A}^{w}$ is the weighted adjacency matrix of $\mathcal{E}$. Each product receives reviews and maintains a review list $R_i=\{ r_{i,1}, \dots, r_{i,|R_{i}|}\}$. 
Let $\mathbf{v}_{i,k}$ be the review-aspect association vector of $r_{i,k}$ and $V^{l}_{i}=\{v^{l}_{i,k}\}_{k=1}^{|R_{i}|}$ be the collected $A_l$'s association list of $P_i$. The product-aspect association vector $\mathbf{u}_{i} \in \mathbb{R}^{|L|}$ can then be obtained by aggregating all the review-aspect association vectors in the review list through $Agg(.)$, a permutation-invariant aggregation function: 
\setlength{\abovecaptionskip}{-0.005cm}
\setlength{\belowcaptionskip}{-0.000cm}
    
\begin{table*}[!ht]
    \centering
    \caption{Three review-aspect association examples are presented to show the different difficulty levels mining these aspects. The mining difficulty for an aspect depends on its relationship to the related corpus (More challenging from top to bottom).  }
    \scalebox{0.72}{
    \begin{tabular}{l|l|l|l|l}
    \hline
       Relation with corpus &  Product type & Aspect & Related corpus & Raw review texts \\
        \hline
        Exact match & Table & Very sturdy & very sturdy & This table is \textcolor{red}{very sturdy}.\\
        \hline
        Synonym & Candle & Looks great & looks perfect & I bought this product for my wife and it \textcolor{red}{looks perfect}  for me. \\
        \hline
        Summary &Carpet cleaning &Great carpet machine&did a great job cleaner cleaning my carpet &This Compact Carpet Cleaner \textcolor{red}{did a great job cleaning my carpet }! \\
    \hline
    \end{tabular}
    }
    \label{tab:example_ra_asso}
    \vspace{-4mm}
\end{table*}
\begin{equation}
    \mathbf{u}_{i} = [Agg(V_i^{1});\dots;Agg(V_i^{|L|})].
    \label{eqn:agg_u}
\end{equation}
The binarized top-$K_l$ aspect label vector $\mathbf{y}^{K_l}_{i}=\mathbf{y}_{i}$ can then be constructed by ranking $\mathbf{u}_i$ and picking up the top-$K_l$ product-aspect associations of $P_i$ to be positive. Let $\mathbf{x}_{i} \in \mathbb{R}^{d_{P}}$ be the node feature of $P_i$, the product aspect mining task is using $\mathbf{x}_i$ and graphical information in $\mathcal{G}$ such as $\mathbf{A}^w$ to predict $\mathbf{y}_i$. Note that the number of positive labels can be less than $K_l$, resulting in the multi-label nature of $\mathbf{y}_i$.



\begin{table}[ht]
    \centering
    \caption{Head and tail aspect examples with frequency information over a three-year period.}
    \scalebox{0.73}{\begin{tabular}{c|c|c|c}
    \hline
        Head aspect &  3-year frequency & Tail aspect & 3-year frequency \\
        \hline
        Looks great	& 1,955,762 & Buttons are good quality & 145\\
Works great	&1,796,908	&Great fit and finish&	145	\\	
Very sturdy	&1,199,984 &Simple plug and play&	145\\	
    \hline
    \end{tabular}}
    \label{tab:aspect_count}
    \vspace{-4mm}
\end{table}
\subsubsection{Predicting future aspect trends for products} 
In practice, products are released at different times and start receiving reviews over time. Consider in a discrete time period $\mathcal{T} = \{0, \dots, T\}$, the product network is evolving and $\mathcal{G}_t=\{\mathcal{P}_t, \mathcal{E}_t\}$ is the product network snapshot at time $t \in \mathcal{T}$. $P_i \in \mathcal{P}_{t>t^{start}_i}$ only occurs after its proposed time $t^{start}_{i}$ in the monotonically increasing product node list $\mathcal{P}_t$ over time. The product similarities edge set $\mathcal{E}_t$ and its weighted adjacency matrix $\mathbf{A}^w_t$are also changing as $t$ changes.
Products begin receiving reviews once become available to customers, therefore the corresponding review lists $R_i(t; t^{start}_{i}) = R_{i,t}$ are also expanding over time. 
Any review $r_{i,k}$ in $R_{i,t}=\{ r_{i,1}, \dots, r_{i,|R_{i,t}|}\}$ also has a proposed time $t^{prop}_{i,k}, t_i^{start} \leq t^{prop}_{i,k} \leq t$, For simplicity, assume $R_{i,t}$ is ordered by ascending review proposed time.
Now $V^l_{i}$ will be $V^l_{i,t}$ with reviews up to $R_{i,t}$, and the product-aspect association vector becomes $\mathbf{u}_{i,t}$ by replacing the $V_i^l$ to $V^l_{i,t}$ in Eqn (\ref{eqn:agg_u}).


Let the above $t$ be the present time, and $\mathbf{x}_{i,t}$ be the node feature of $P_i$ in $\mathcal{G}_t$. Suppose at a future time $t'=t+\Delta t$, the product aspect vector and the top-$K_l$ aspect label vector is $\mathbf{u}_{i, t'}$ and $\mathbf{y}_{i,t'}$.
The $\Delta t$-future product aspect forecasting problem at time $t$ is to predict $\mathbf{y}_{i,t'}$ using $\mathbf{x}_{i,t}$ and $\mathcal{G}_t$. 
Moreover, we are interested in forecasting top product aspects $\mathbf{y}_{i,t+\Delta t_n}$ at multiple future with $N_{\Delta_T}$ different time lengths $\Delta T \triangleq \{\Delta t_n\}^{N_{\Delta_T}}_{n=1}$ simultaneously using information till $t $. In practice, the accessible information is till the present $t$ (referred to as `now'). Yet setting `$t$ at now' to an earlier stage will therefore meet with the old product that is earlier `new', together with a smaller product network that has fewer products with shorter review lists.

\subsection{Motivations and challenges of the problem}
\label{sec:challenge}
\subsubsection{Motivation}
We illustrate the motivation of the product aspect forecasting problem. 
In a large e-commerce system, existing products keep receiving reviews from buyers, and sellers continue to release more and more new products (Fig. \ref{fig:split_figure1}). The high-level target of this problem is predicting the customer feedback for products in the long future: e.g., \textit{what will be the top aspects of a new product six months or three years later}? This problem might be easier to answer for products with long histories - they have many customer reviews, thus simply extracting all aspects till now and counting the most frequent ones can already obtain statistically significantly satisfying results. However, for the larger and more important group of newly released products with limited reviews, such a strategy will give a much less reliable aspect forecasting, as aspects that currently have low frequency might be frequently mentioned in future reviews. On one hand, customers cannot trust products with less accurate aspects prediction; on the other hand, sellers worry about long-history products monopolizing the market. Therefore, a robust and reliable solution clearly relieves the dilemma that new products have limited reviews to help forecast on both the seller and customer side despite the non-trivial problem nature. 

\subsubsection{Challenges}
Forecasting aspects for `new' products can not be effectively achieved by trivially aggregating mined results from aspect mining methods, as the number of reviews is limited besides their subjectiveness. 
Also, the scale of reviews can be very large and aspect list length $L$ can also be large in practice, leading to high sparsity of positive associations and extremely imbalanced aspect frequencies (Table \ref{tab:aspect_count}). Therefore the top-K aspect retrieving formulation also introduces association sparsity and extremely imbalanced aspect frequencies, making it more challenging.

Another challenge is that the `new' products are distributed at different time points, resulting in unaligned review list series evolving processes. It is because reviews will keep occurring for products once they are released, and all the products will receive different reviews at different times. Since forecasting aspect trends for the `new' products with limited reviews is more important and challenging, trivially training a graph neural network from a single product graph is also not effective. 
The evolving product-review associations and product similarities need to be efficiently extracted, aggregated, and embedded across the whole time range to avoid over-fitting effects.

The third challenge is that the aspect evolving trends will be quite different in shorter and longer time periods, as $\mathbf{y}_{i,t'}$ will reflect the genuine converged customer feedback when $\Delta t \rightarrow \infty$ and the recent popular thoughts from customers when $\Delta t$ is small. 
Therefore predicting top aspect trends for products at different $\Delta t$ future indicates capturing different stages of product evolving from the entire `new' (with limited reviews) to `old' (with adequate reviews) evolving process from time to time. Trivially using one single product embedding at current time $t$ without further future-length aware adjustment will result in unsatisfactory performance. 



%% file: main_paper/methods.tex
\section{Methods}
\begin{figure*}[ht]
\centering
 \includegraphics[width=\linewidth]{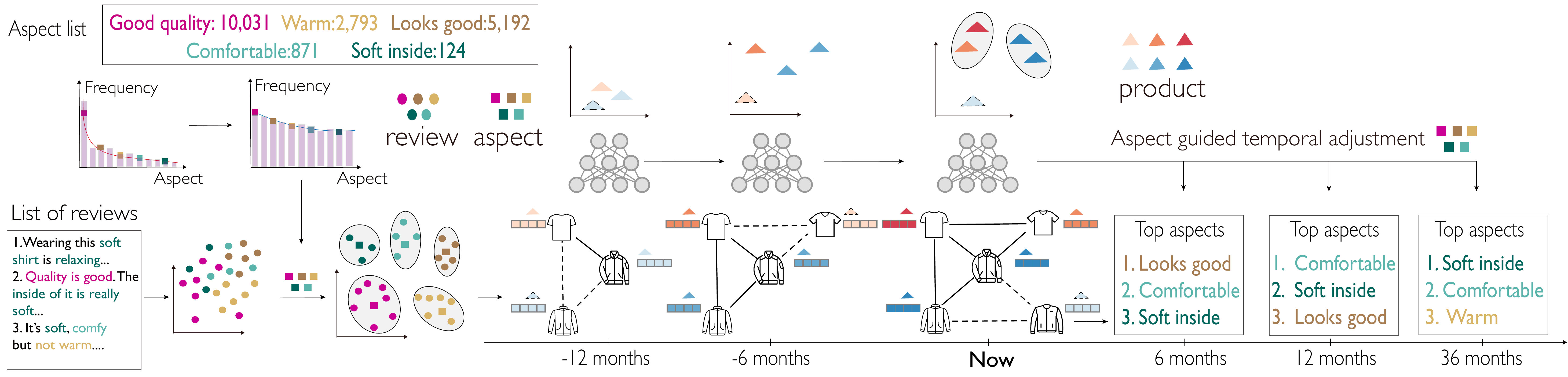}
\caption{ForeSeer first resamples review-aspect associations and co-embed aspects and reviews. It then progressively embeds products leveraging temporal product graphs. Finally, it exploits temporal information to adjust product and aspect embeddings to provide multi-time future forecasting.}
\label{fig:figure2}
\vspace{-4mm}
\end{figure*}
We propose ForeSeer, a textual mining and product embedding framework progressively trained on temporal product graph to predict the dynamic product embeddings at a future time point $t+\Delta t$ using all information till the current time point $t$. 
ForeSeer then exploits the product embeddings to predict the top aspects of that future time. 
ForeSeer is a three-stage framework (Fig. \ref{fig:figure2}). 
In the first stage, we exploit contrastive learning to co-embed reviews and aspects, leveraging the aspect semantic information to acquire better review embeddings. 
We resample review-aspect associations to increase the presence of rare aspects so that the review embeddings can be less biased to frequently-appeared aspects. In the second stage,
we progressively train the product graph embedding network on the temporal product graph. This helps ForeSeer see more new products and the new reviews they received, as well as evolving product similarity information, resulting in an efficient product embedding alignment.
In the third stage, we develop a novel Product-Aspect Temporal Attention (PATA) module to further adjust product embeddings with the help of learned aspect embeddings. PATA module helps to forecast future product-aspect associations with explicit consideration of temporal information, thus providing time-sensitive forecasting.

\vspace{-2mm}
\subsection{A base model}
\label{sec:base_model}
We first introduce a base model that extends aspect mining methods for aspect forecasting and later contrast this base model with Foreseer to clarify the key technical ideas of ForSeer. 
Let $\mathbf{R}= \{r_j\}^{|\mathbf{R}|}_{j=1}$ be the list of reviews collected from all products. The base model builds a BERT-based textual encoder that is pretrained on all reviews with the mask language modeling (MLM) objective \cite{devlin2018bert}. The model can then be used to embed $r_j$ as $\mathbf{h}_{j} = \text{BERT}(r_{j})$. 
At present time $t$ (`now'), the learned review embedding $\mathbf{h}_j$ can then be aggregated to the product $P_i$ they belong to as $\mathbf{f}_{i}$ via an aggregator function. Let $ R_i$ be the review list of $P_i$ at time $t$ and $r_{i,k} \in R_i$ be one of the review of $P_i$, the aggregation process is:
\begin{equation}
    \begin{aligned}
    \mathbf{f}_{i} = Agg(\{\mathbf{h}_{i,k} | r_{i,k} \in R_i\}).
    \end{aligned}
    \label{eqn:agg_f}
\end{equation}

The base model then leverages $\mathbf{x}_i= [\mathbf{f}_{i}; \textbf{SE}_i]$ as a node feature of $P_i$ and trains a graph neural network on the last observed product graph $\mathcal{G}_{last}$ to further get a product embedding $Z_{last}$. Here, $\textbf{SE}_i$ is the static product-specific embedding that encodes other product-related information, such as seller descriptions, which will not change over time. Let $\mathbf{A}_{last}^{w}$ be the weighted adjacency matrix of $\mathcal{G}_{last}$, and $X$ be the product node feature matrix of all products that make up $\mathbf{x}_i$, where $P_i \in \mathcal{G}_{last}$. Let $\sigma$ be an activation function and $H^{(0)}= X$, for $H^{(l_p)}$, the output of the $l_p$-th GNN layer we have: 
\begin{equation}
    \label{eqn:base_gnn}
    \begin{aligned}
    H^{(l_p)}= \sigma([\mathbf{A}^{w}_{last}H^{(l_p-1)}W^{l_p}]).
    \end{aligned}
\end{equation}

Let $Z_{last}=H^{(L_p)}$ be the resulting embedding matrix of all products after the final layer, its $i$-th row $\mathbf{z}_i$ is the embedding of $P_i$ and is used to get the prediction of top aspects in the future. 
Suppose we are interested in multiple futures $\Delta T\triangleq\{\Delta t_n\}^{N_{\Delta T}}_{n=1}$, 
we use $N_{\Delta T}$ classifier heads $\{\text{CLS}_{P \rightarrow A, n}\}^{N_{\Delta T}}_{n=1}$ to map $Z_{last}$ and different $\Delta t_n$ to the prediction $\hat{\mathbf{y}}_{i,t+\Delta t_n}$ for different future $t + \Delta t_n$. They are finally optimized with a multi-time multi-label cross-entropy objective:
\begin{equation}
    \setlength{\belowdisplayskip}{-0.001mm}
    \begin{aligned}
      \hat{\textbf{y}}_{i, t+\Delta t_n} &= \text{CLS}_{P \rightarrow A, n}(\mathbf{z}_i, \Delta t_n ), \\
    \mathcal{L}^{(i)}_{P\rightarrow A} &= \frac{1}{N_{\Delta T}\cdot |\mathbf{L}|}\sum^{N_{\Delta T}}_{n=1}\sum^{|L|}_{l=1} -y^{l}_{i, t+\Delta t_n}\log \hat{y}^{l}_{i, t+\Delta t_n} .  
    \end{aligned}
    \label{eqn:pa_cls}
\end{equation}


\vspace{-4mm}
\subsection{Mining aspect to enhance review embedding} 
\label{sec:mine_aspect}


The major limitation of the base model is it only pre-trains the language model using review texts instead of clearly capturing the review-aspect association. Because reviews are usually long and subjective (Table \ref{tab:example_ra_asso}), it is very inefficient to capture signals of potential aspects from the embedding of the entire review text. Also, to have more descriptive and diverse aspects, the aspect list can be very long. Under this situation, the positive aspect association in review texts will be very sparse, and the aspect distribution can be really imbalanced and long-tailed.
We thus propose to refine the review embedding by fine-tuning the review encoder to mine aspect associations. The key intuition is to embed the review focusing on representing sparse aspect associations. 

Specifically, given a review text $r_j$, we fine-tune its embedding $\mathbf{h}_j$ to capture the review-aspect association vector $\hat{\textbf{v}}_{j}$ with a classifier head $\text{CLS}_{r\rightarrow a}(.)$ and a multi-label cross-entropy loss objective:
\begin{equation}
    \setlength{\abovedisplayskip}{-0.001mm}
    \setlength{\belowdisplayskip}{-0.001mm}
    \begin{aligned}
     \hat{\textbf{v}}_{j} = \text{CLS}_{r\rightarrow a}( \mathbf{h}_{j}); \ 
    \mathcal{L}^{(j)}_{cls,r\rightarrow a} = \frac{1}{|\mathbf{L}|}\sum^{|\mathbf{L}|}_{l=1}- v^{l}_{j}\log\hat{v}^{l}_{j}.  
    \end{aligned}
    \label{eqn:ra_cls}
\end{equation}

Note that here the association between the review texts and the resulted aspects will not change with time; it is therefore compatible to combine this objective with the MLM objectives \cite{devlin2018bert} in the BERT pre-training stage.
The resulting prediction of the association of the review aspect $\hat{\mathbf{v}}$ can be viewed as part of the refined review embeddings. For instance, the updated representation $\mathbf{h}'_{j}=[\mathbf{h}_j;\hat{\mathbf{v}}_j]$ can be the concatenation of BERT embedding and predicted association vectors. Having it can thus better capture sparse aspect association information during review aggregation using Eqn. (\ref{eqn:agg_f}). 

\subsection{Aspect-guided review embedding matching} 
However, the limitation of the above refinement is that it has no idea that some aspects might describe similar meanings. It thus lacks the ability to efficiently capture the tail aspects that might have a similar meaning to the top frequent aspects but are rare and more diverse.
Incorporating class names is useful when classes have their own description \cite{meng2020text}. Therefore, our next key improvement is to jointly embed reviews and aspects through cross-domain contrastive learning. 
The key intuition is that by also embedding aspects, tail aspects can be closer to head popular aspects if they have similar semantic meanings. The resulting aspect embeddings can therefore guide two review embeddings to be closer if they point to aspects with similar meanings. Moreover, if we treat the base model as a review feature extractor, training a cascaded contrastive learner will enable an effective way to train multiple review feature extractors and incorporate resampling and ensemble techniques.  

Specifically, we train multiple feature extractors by instantiating $M$ base aspect extraction replicas with different reweighting factors $\alpha_1, ..., \alpha_M$ in addition to the standard aspect extraction instance ($\alpha_0=0$). We use $\textbf{h}^{(\alpha_m)}_{j}$ and $\hat{\textbf{v}}_j^{(\alpha_m)}$ to denote the trained features output by the $m$-th replica follows Eqn (\ref{eqn:ra_cls}). We then downsample frequent aspects to encourage the model to focus on review-aspect associations from infrequent aspects. For replica $m$, we first construct a bin for each aspect by assigning reviews that have positive associations with that aspect. Here, a review might be allocated to multiple aspect bins. We thus apply the resampling strategy hierarchically, which chooses the $l$-th aspect bin with probability $p_{l}$ by reweighting and then normalizing the importance of the aspect based on the reciprocal of the $\alpha_m$-reweighted aspect frequency: 
\vspace{-1mm}
\begin{equation}
    \begin{aligned}
    p_{l} = \frac{w_{l}^{(m)}}{\sum_l w_{l}^{(m)}}, \ w_{l}^{(m)} = (\frac{1}{freq(A_l)})^{\alpha_m} , \ 
     \alpha_m \geq 0.          
    \end{aligned}
\end{equation}
\vspace{-1mm}
The reweight factor $\alpha_m$ leads the model to focus more on the aspects of tail (large $\alpha_m$) or head (small $\alpha_m$). $\alpha_m = 0$ corresponds to the standard aspect extractor that conduct uniform sampling. Then, a review $r_j$ is sampled from the chosen aspect bin with an assigned importance weight based on the number of review-aspect associations and the reweighted aspect importance. 

This resampling strategy enables the possibility to ensemble the features from all $M$ feature extractors and perform mixture-of-experts (MoE) learning in the contrastive learner. 
We can now exploit contrastive learning to co-embed reviews and aspects so that the embedding of reviews can pay more attention to rare aspects. 
Specifically, we first build review and aspect embedding networks $E_R$ and $E_A$ with the same output dimension $d$. 
$E_R$ encodes the embeddings of the review $r_j$ and the predicted associations from all replicas of the extraction of aspects simultaneously and matches them into a calibrated embedding of the review $\mathbf{z}^{re}_{j}$:
\vspace{-1mm}
\begin{equation}
    \begin{aligned}
    \mathbf{z}^{re}_{j} = E_R(\text{CONCAT}([\textbf{h}_{j}; \hat{\textbf{v}}_{j};\textbf{h}^{(\alpha_1)}_{j};  \hat{\textbf{v}}^{(\alpha_1)}_{j};...;\textbf{h}^{(\alpha_M)}_{j};  \hat{\textbf{v}}^{(\alpha_M)}_{j}]).
    \end{aligned}
\end{equation}

For aspect $A_l$, we use SpanBERT \cite{joshi2020spanbert} to get the pre-trained features and feed it into $E_A$ to get the aspect embedding $\mathbf{z}^{asp}_{ l}$:
\begin{equation}
    \begin{aligned}
    \mathbf{z}^{asp}_{l} = E_A(\text{SpanBERT}(A_l)) .     
    \end{aligned}
\end{equation}

We then utilize the aspect embeddings $\mathbf{z}^{asp}_{l}$ of $A_l$ to guide the optimization of the calibrated review embedding $\mathbf{z}^{re}_{j}$ of review $r_j$ using the inner product of $\mathbf{z}^{re}_{j}$ and $\mathbf{z}^{asp}_{l}$: 
\begin{equation}
    \begin{aligned}
    s^{l}_{j} &= (\mathbf{z}^{re}_{j})^T \mathbf{z}^{asp}_{l}; Z^{asp}=[\mathbf{z}^{asp}_{1};\dots;\mathbf{z}^{asp}_{|L|}]; \\
    \mathbf{s}_j &= (\mathbf{z}_{j}^{re})^TZ^{asp} = [s^{1}_{j};\dots;s^{|L|}_{j}] \in \mathbb{R}^{|L|}.
    \end{aligned}
\end{equation}

We finally employ a multi-label cross-domain contrastive loss similar to NT-Xent loss ~\cite{sohn2016improved} to maximize positive association and minimize negative association denoted by $\mathbf{v}_j$:
\begin{equation}
    \begin{aligned}
    \mathcal{L}_{CL}(\mathbf{s}_j, \mathbf{v}_j;\tau) = -\log \sum_l  \frac{v^{l}_j exp(s^{l}_{j}/\tau)}{\sum_k exp(s^{k}_{j}/\tau)}. 
    \end{aligned}
    \label{eqn:ra_cl}
\end{equation}

By jointly learning an informative, low-dimensional embedding space for reviews and aspects, the resulting $\mathbf{s}_j$ has two advantages. First, it will efficiently ensemble learnings from multiple aspect extraction replicas, thus more effectively capturing tail aspects. Second, by minimizing Eqn (\ref{eqn:ra_cl}), $\mathbf{s}_j$ will encode the ranking information of the aspects for $r_j$. This indicates a more flexible way to retrieve $K_a$ positive associations. By simply tuning $K_a$,  more (recall-inclined) or less (precision-inclined) positive associations can be given, resulting in a more flexible aspect extraction module.



\begin{figure*}[ht]
\setlength{\belowcaptionskip}{-0.01cm}
\centering
 \includegraphics[width=0.81\linewidth]{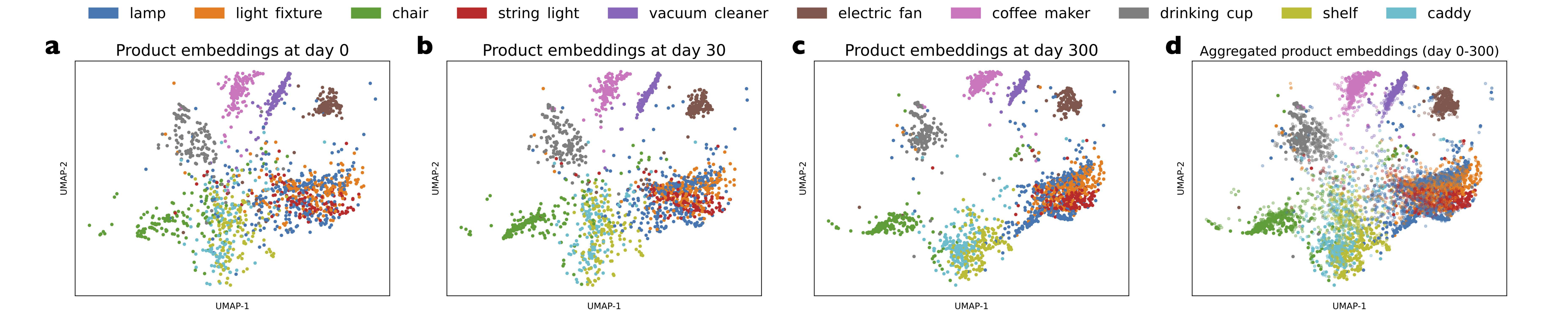}
\caption{UMAP plots visualizing product embeddings at day 0 (\textbf{a}), day 30 (\textbf{b}), day 300 (\textbf{c}), and aggregation of product embeddings at all of the three time points (\textbf{d}).}
\label{fig:UMAP_product}
\vspace{-5mm}
\end{figure*}

\subsection{Progressive temporal graph-based product embedding}



The major limitation of the graph-based product embedding module of the base model is that it is only trained on the last observed product graph snapshot with the product features at that time. In fact, only a small ratio of products are `new' in the last observed graph snapshot. This leads the model to have severe overfitting effects. We thus propose to progressively train the model on the temporal product graph snapshot series rather than only on the last observed graph. The key intuition is that the model can not only see more `new' products during training but also be aware of the other temporal evolving information, including product similarities and how a product becomes from `new' to `old' by receiving more and more reviews. 

Without loss of generality, we now let $t \in \mathcal{T}$ be a time point we want to train our model at and let the graph be the snapshot $\mathcal{G}_t$ at that time. To enable progressive training, multiple fixations are needed. First, the feature vectors of product nodes need to be aware of the evolving temporal information, such as the related information of the progressive product-aspect associations until time $t$. Given a review $r_{i,k} \in R_{i,t}$, the review list of $P_i$ at time $t$, we aggregate the learnt embedding $\mathbf{z}_{i,k}$ and the predicted review-aspect association $\mathbf{s}_{i,k}$ for all reviews in $R_{i,t}$ to acquire the product feature $\textbf{f}_{i,t}$ of $P_i$ at time $t$:
\begin{equation}
    \begin{aligned}
    \textbf{f}_{i,t} = Agg(\{\text{CONCAT}([\mathbf{z}_{ i,k}, \mathbf{s}_{i,k}]) | r_{i,k} \in R_{i,t}\}).
    \end{aligned}
\end{equation}

We then incorporate the related temporal information of $P_i$ at time $t$. Similar to the positional embedding \cite{vaswani2017attention}, the 2d-dimensional temporal embedding of a discrete-time index $t_0$ is:
\begin{equation}
    \begin{aligned}
    Emb_{t_0} = [\sin(\omega_1 t_0), \cos(\omega_1 t_0), \dots, \sin(\omega_{d} t_0), \cos(\omega_{d} t_0)].
    \end{aligned}
\end{equation}

For $P_i$ at time $t$, there are three pieces of temporal information that need to be incorporated: its proposed time $t_i^{start}$, the training time $t$ and the gap between these two times $t_i^{gap}=t - t_i^{start}$. The resulting temporal embedding therefore has three sub-parts:
\begin{equation}
    \begin{aligned}
    T^{emb}_{i,t} = \text{CONCAT}([ Emb_{t_i^{start}}, Emb_t, Emb_{t^{gap}_i}]).
    \end{aligned}
\end{equation}

We then compose the overall node feature vector $\mathbf{x}_{i,t}$ for $P_i$ at training time $t$ with three parts of features mentioned before: 
\begin{equation}
    \begin{aligned}
    \mathbf{x}_{i,t} = \text{CONCAT}([\mathbf{f}_{i,t}; T^{emb}_{i,t}; \textbf{SE}_i]).
    \end{aligned}
\end{equation}

Let $\mathbf{A}^w_t$ be the normalized weighted adjacency matrix of $\mathcal{G}_t$, to enable progressive training, we sequentially train the graph neural network at a subset of graph snapshots $\{\mathcal{G}_{t_1}, \dots, \mathcal{G}_{t_Q} | t_q \in \mathcal{T}\}$. We further refine the GNN layer that contains an extra multi-layer perceptron (MLP) module that focuses on learning product-specific information. 
Specifically, let $W^{l}_{t_{q-1}}$ and $W^{l}_{MLP, t_{q-1}}$ be the weight matrix of $l$-th GNN layer and the extra MLP module after training the network on $\mathcal{G}_{t_{q-1}}$ and let $H^{(0)}_{t_{q}}=\mathbf{x}_{i,t_{q}}$, the product embedding process in Eqn (\ref{eqn:base_gnn}) will be updated at $t_{q}$ as:
\begin{equation}
    \begin{aligned}
    H^{(l_p)}_{t_{q}}= \sigma(\text{CONCAT}([\mathbf{A}^{w}_{t_{q}}H^{(l_{p-1})}_{t_{q}}W^{l_p}_{t_{q-1}};
    W^{l_p}_{MLP,t_{q-1}}H^{(l_{p-1})}_{t_{q}}])).
    \end{aligned}
\end{equation}

At present time $t$, we take $Z^{P}_t = H^{(L_p)}_{t}$ got on $\mathcal{G}_t$ as the resulting product embedding matrix. By progressively updating the network, ForeSeer can see more new products and how they `grow' from `new' to `old', which is helpful in efficiently aligning the product embedding.

\subsection{PATA: Aspect-guided product embedding adjustment for multi-future forecasting}



One limitation of the base model is it uses only one product embedding for all $\Delta t_n$ future predictions without any refinement when capturing future product-aspect associations.
We thus develop a novel multi-head product-aspect temporal attention (PATA) module to adjust product embedding based on learned aspect embedding $Z^{asp}$. Our intuition is to adjust product embedding with a designated attention module guided by aspect embedding. This novel design empowers ForeSeer with the ability to predict both short and long-future time ranges with lower computational costs.

Specifically, given $N_{\Delta T}$ target futures, for each $\Delta t_n$, we set up a query network $EP^{Q}_{\Delta t_n}$ and a key network $EA^{K}_{\Delta t_n}$ to get the product query matrix from $Z_t^{P}$ and aspect key matrix from $Z^{asp}$:
\begin{equation}
    \begin{aligned}
    Q^{P}_{t+\Delta t_n} &= EP^{Q}_{\Delta t_n}(Z^{P}_{t}).\\
    K^{asp}_{t+\Delta t_n} &= EA^{K}_{\Delta t_n}(Z^{asp})
    \end{aligned}
\end{equation}

We then calculate cross-attention score $Attn_{t+\Delta t_n}$ between product query and aspect key for $\Delta t_n$ future:
\begin{equation}
    \begin{aligned}
    Attn_{t+\Delta t_n}= Softmax(\frac{Q^{P}_{t+\Delta t_n}(K^{asp}_{t+\Delta t_n})^T}{\sqrt{d}}) \in \mathbb{R}^{|\mathcal{P}_t| \times |L|}.
    \end{aligned}
\end{equation}
We then set up a shared value network $EV^{val}$ that takes $Z^{P}_{t}$ as input and outputs value matrix $Val^{P}_{t}$.
Let $val_{i,t}$ and $attn_{i, t+ \Delta t_n}$, the $i$-th row of $Val^{P}_{t}$ and $Attn_{t+\Delta t_n}$, be the value and attention score vector for $P_i$, we obtain the final product embedding for $\Delta t_n$ future as:
\begin{equation}
    \begin{aligned}
    \mathbf{z}^{final}_{i, t+ \Delta t_n} &= attn_{i, t+\Delta t_n} \circ val_{i,t}. 
    \end{aligned}
\end{equation}
We next exploit the final product embedding to get the final prediction using Eqn (\ref{eqn:pa_cls}). With the above design, we can acquire different product embedding for different $\Delta t_n$ future prediction tasks. Using a shared value network is especially efficient when $N_{\Delta_T}$ is large and $d \ll |L|$. This also regularizes the model by sharing weights for learning multiple imbalanced top-$K_l$ aspect retrieving tasks. 

%% file: main_paper/experiments.tex
\section{Experiment}
In this section, we evaluated ForeSeer on a large scale e-commerce dataset with a variety of tasks, aiming to answer the following research questions (RQs):
\begin{itemize}
    \item \textbf{RQ1:} How does ForeSeer perform on product aspect forecasting without or with genuine annotated aspects for reviews? 
    \item \textbf{RQ2:} Is ForeSeer sensitive to aspects with long trends?
    \item \textbf{RQ3:} How does different components contributes to the success of ForeSeer on capturing review-aspect associations?
    \item \textbf{RQ4:} Is the PATA module in ForeSeer able to capture temporal information for different lengths of future?   
    \item \textbf{RQ5:} How does the product embedding captured by ForeSeer evolve over time?   
\end{itemize}

\begin{table*}[t]
    \centering
    \caption{Performance of ForeSeer and comparison approaches on aspect forecasting at three different time gaps (3 months, 6 months, 3 years) under settings of with annotated aspects and without annotated aspects.}
    \scalebox{0.82}{
    \begin{tabular}{l|l|cccc | cccc}
    \hline
         \multirow{2}{*}{$\Delta t$} & \multirow{2}{*}{Method}  &\multicolumn{4}{c|}{Without annotated aspects} &\multicolumn{4}{c}{With annotated aspects} \\
         &   &AUPRC & AUROC & Kappa & Max F1 & AUPRC & AUROC & Kappa & Max F1 \\
         \hline
        \multirow{6}{*}{3 months} & Frequency baseline & 21.0 ±5.7 & 84.8±4.1 & 28.6 ±6.2 & 37.4 ±5.6 & 91.2 ±4.2& 96.9± 2.0 & 86 ±5.4 & 90.0± 4.2 \\
         &  MLP  & 13.1± 7.2 &89.8± 3.2 & 4.5 ±4.3 & 25.1± 6.5 & 14.6± 7.7 & 91.3 ±2.9 & 4.9 ±4.1 & 28.5 ±7.1   \\     
         &  LSTM   &17.0 ±6.0 & 93.7± 2.5 & 6.5 ±5.1 & 29.6 ±6.1 & 17.4 ±6.1& 93.8 ±2.6 & 7.3 ±5.5 & 30.1 ±6.2 \\ 
         &  GRU  & 17.3± 6.2 &93.6± 2.6 & 7.0 ±5.5 & 30.1± 5.5 & 18.2± 6.2 & 94.0 ±2.6 & 5.9 ±5.3 & 31.0 ±6.3   \\ 
         & GNN & 19.2 ± 6.6 & 95.1 ± 3.8 & 25.1± 8.1& 32.3± 6.8& 20.9 ± 6.5 & 97.2± 3.9 & 24.9± 7.0& 34.2 ±6.6  \\
        & ForeSeer (Ours) &\textbf{84.1 ±7.2} & \textbf{99.9± 0.2} & \textbf{75.8± 7.9}& \textbf{83.9 ±5.9} & 80.2 ± 8.4 & 99.9 ±0.1 & 72.3 ±8.2 & 80.0± 7.2\\
         \hline
        \multirow{6}{*}{6 months} & Frequency baseline & 20.2± 5.6 & 84.3± 4.2 & 27.5± 6.2 & 36.1± 5.6& 87.4 ±5.1& 96.4 ±2.2 & 81.8 ±6.0 & 86.3 ±4.9 \\ 
         &  MLP  & 11.1± 7.1 & 88.4± 3.5 & 4.3 ±4.1 & 23.5± 6.7 & 13.9± 7.5 & 90.9 ±2.7 & 5.1 ±4.0 & 27.6 ±6.8   \\ 
         &  LSTM   & 14.9± 5.7 &93± 2.8 & 6.6 ±5.2 & 27.5± 6.0 & 15.3 ±5.8 & 93.0 ±2.9 & 7.2 ±5.5 & 27.8± 6.1 \\        
         &  GRU  &  15.4± 5.9  & 92.9 ±2.8 & 6.7 ±5.3 & 28.1± 6.2 & 16.2± 6.0 &93.3± 5.6 & 28.9 ±6.3& 26.3 ±5.3\\   & GNN & 18.1 ± 6.5 & 93.8 ± 3.5 & 23.0± 7.7& 29.4± 6.7& 19.3 ± 6.4 & 95.9± 3.3 & 22.9± 6.9& 31.9 ±6.6  \\  
        & ForeSeer (Ours)  &\textbf{81.8± 7.7} & \textbf{99.8± 0.2} & \textbf{74.2± 8.1}& \textbf{82.2 ±6.3}& 82.3±8.6 & 99.7 ±0.3 & 73.6 ±8.5 & 82.6±6.9\\
         \hline
        \multirow{7}{*}{3 years} & Frequency baseline & 18.6 ±5.4 & 83.8 ±4.2 & 25.4± 6.1 & 34.0± 5.6 & 81.0± 6.1 & 96.2± 2.3 & \textbf{75.8 ±6.5} & 81.2± 5.3 \\ 
         &  MLP   &  10.2± 4.9 & 86.3 ±3.0 & 3.1± 3.0 & 21.2± 5.3& 11.7± 5.3 &89.7 ±4.2& 5.5± 5.2 &23.3± 6.1  \\  
         &  LSTM   &  11.4± 5.0 & 90.3 ±3.3 & 5.8± 4.9 & 23.4± 5.7& 12.0± 5.2 &90.3 ±3.5& 6.6± 5.3 &24.1± 6.0  \\        
         &  GRU  & 12.1 ±5.2 & 90.3 ±3.3 & 6.0 ±5.1 & 24.4 ±5.9 & 13.1± 5.4& 90.7±6.7 & 5.9±5.1 & 25.1±6.0 \\  
         & OA-Mine & 5.9 ± 3.9 & 78.6 ± 5.9  & 3.4 ± 2.7 & 6.1 ± 2.6& 5.9 ± 3.9 & 78.6 ± 5.9  & 3.4 ± 2.7 & 6.1 ± 2.6  \\
         & GNN & 16.1 ± 6.3 & 92.1 ± 3.1 & 20.9± 7.1& 28.6± 6.6& 17.3 ± 6.4 & 93.0± 2.7 & 20.6± 6.9& 29.7 ±6.5  \\
        & ForeSeer (Ours) &\textbf{68.5 ± 8.9} & \textbf{98.9± 1.1} & \textbf{65.6 ± 8.5}& \textbf{73.2 ± 6.9}& \textbf{83.6 ±7.9} & \textbf{99.9± 0.1} & 74.7 ± 7.7 & \textbf{82.8± 6.4} \\
    \hline
    \end{tabular}
    }
    \label{tab:main}
    \vspace{-3mm}
\end{table*}
\vspace{-3mm}
\subsection{Experimental setup}
\subsubsection{Dataset}
We evaluated our ForeSeer framework on a large-scale e-commerce dataset that contains a dynamic product network with 11,536,382 time-stamped review events in a three-year period. The product network is a series of homogeneous dynamic product similarity graphs with 1,096 daily time-stamped snapshots. The final graph has 11,000 product nodes from 418 different product types based on their name, high-level property, and usage, such as "Lamp", "TABLE", and "Caddy". The unpublished product nodes will not be added to the graph until their proposed times are reached. The average start date of products is 77. 
The first snapshot has 999,614 edges and the final snapshot has 2,400,404 edges. Each edge represents a similarity ranging from 0 and 1 between the two products. We calculate the similarities based on multiple factors such as user click information. We obtained an aspect list with 10,000 popular aspects extracted via sequential extraction tools \cite{yan2021adatag, yang2022mave} and 30,217,638 review-aspect associations from these reviews. We constructed the top-$K_l$ ($K_l = 10$) aspects label for every product node at all timesteps by aggregating all the review-aspect associations it has and picking up the top-ranked aspects after normalization. 
Example head and tail aspects with their frequency over three years are shown in Table \ref{tab:aspect_count}. While the top 3 aspects have more than 1,000,000 counts, the tail aspects only have 150 counts from more than 11 million review texts, demonstrating extreme aspect imbalances. 



\vspace{-2mm}
\subsubsection{Tasks}
We studied ForeSeer with three challenging tasks: single review-aspect association prediction, multi-time aspect forecasting, and multi-time link prediction. 

\noindent \textbf{Single review-aspect association prediction:} We model the single review-aspect association prediction as a multi-label classification problem. We split 80\% of the reviews as the training set and leave the remaining 20\% as the test set. 

\noindent \textbf{Multi-time aspect forecasting:} To further assess the effectiveness of the learned review associations, we tested the multi-time aspect forecasting task with two settings: (i) ground truth association accessible, (ii) only learned approximate association accessible. For this task, we chose to predict $\Delta t=3,6$ months and 3 years (always predicting the last observed label) at every test timestep $t$. We use the label at the final time if the target future time is out of range. We aimed to test the inductive performance by randomly splitting 10\% nodes out as the test set.

\noindent \textbf{Multi-time link prediction:} For multi-time link prediction, we chose to predict the link at the current time, 180 days later, and at the final time at every test time $t$. The positive edge ratio is 50\%. 

\vspace{-1mm}
\subsubsection{Comparison approaches}
There is no existing framework that directly handles the multi-time aspect forecasting problem. We therefore designed and compared our method with four types of baselines. \textbf{Frequency baseline} directly aggregates the association learnt in section \ref{sec:mine_aspect} at time $t$ and uses it as the prediction. It can not incorporate other information and the performance will highly depend on the quality of learned aspect associations from reviews. We aim to assess the effectiveness of the multi-time forecasting PATA module.
\textbf{MLP baseline} directly predicts the multi-time future with product features without graphical information. \textbf{Recurrent-based baselines (GRU, LSTM)} predict multi-time future step-by-step by using an autoregressive network structure and also ignore graphical similarity information. We thus aim to assess the importance of using graphical information. \textbf{Graph-based baseline (GNN)} leverages only the final graphical information $\mathcal{G}_{last}$.  
We aim to assess the effectiveness of our progressive training process in multiple $\Delta t$-future. \textbf{Weakly supervised baseline (OA-Mine)} directly mines candidate aspects from reviews and aggregates the candidates as the final predictions. We aim to assess the effectiveness of learned review-aspect associations and the importance of using learning-based models as this baseline does not leverage review-aspect association information.

For multi-time link prediction tasks for different $\Delta t$-future, we only compare our method with graph-based baselines, since other baselines can not handle this task. We built an extra lightweight GNN for multi-task link prediction for our methods and a GNN baseline with the same network structure. For our method, we used features learned from our multi-head product aspect temporal attention module for different futures as the input for different future time predictions. For the graph-based baseline, we used the same features for different future time predictions. 
\begin{table*}[ht]
    \centering
    \caption{Case study showing the aspected we predicted for skin moisturizer. Our method identified time-sensitive phrases (red) while the frequency baseline failed to identify them.}
    \scalebox{0.78}{
    \begin{tabular}{l|l}
    \hline
        Product type & Skin moisturizer \\
    \hline
Count baseline prediction & Smooth, Feels great,  Product is great, Clean, Light weight, Soothing, Love this product, Dries quickly,\\
with one month reviews &  Works great, Great price\\
\hline
ForeSeer prediction with& Smooth, Clean, Great price, Product is great, Soothing, \textcolor{red}{A little goes a long way}, \textcolor{red}{Sensitive skin}, Light  \\
one month reviews& weight, \textcolor{red}{Not greasy}, Good quality\\
\hline
\multirow{2}{*}{Final Label (Ground truth)}& Smooth, Love the texture, Light weight, Product is great, Soothing, Great price, \textcolor{red}{A little goes a long way},\\
& \textcolor{red}{Sensitive skin }, Fragrance free, \textcolor{red}{Not greasy}\\
    \hline
    \end{tabular}
    }
    \label{tab:case_study_1}
    \vspace{-4mm}
\end{table*}
\vspace{-2mm}
\subsubsection{Implementation details}
For the aggregation function $Agg(.)$ used to construct future aspect labels, we used $sum(.)$ with per-aspect z-score normalization operation. Specifically, for each aspect, we got the sum of its occurred association for every product at the end of the time period and calculated the mean and standard deviation of them. We then performed z-score normalization on the sum of the associations of this aspect for every product at time $t$. 
We used $mean(.)$ as the aggregator to get the aggregated feature $f_{i,t}$ of product $P_i$ at time $t$. We maintained both the predicted logits and the binarized association prediction from $\mathbf{s}_{i,k}$. We incorporated a pre-trained product embedding obtained from a multi-modality neural model that encodes information from the product title and descriptions to construct the auxiliary product embedding $\mathbf{SE}_i$.

For simplicity, we only trained one resampling replica with $\alpha=0.5$ and match its embedding with embedding learned by standard aspect extraction replica ($\alpha=0$). 
We used AdamW as the optimizer and $10^{-4}$ as the learning rate to train the aspect-guided embedding match model for 10 epochs on four 16GB Nvidia-V100 GPUs and set the number of retrieved positive associations $K_a=5$ for better precision performance. We used a two-layer MLP with ReLU for all the embedding networks and a linear projection layer for all query and key blocks. All the embedding, query, and key vector output dimensions are 100. To avoid the over-smoothing problem, we used a GNN with one designed graph neural layer and two cascaded MLP layers. We used the same network structure for the GNN baseline. For the MLP baseline, we used a two-layer MLP with a hidden size of 128, for GRU and LSTM baseline we set the number of the layer as 2 with a hidden size of 128. For the lightweight GNN for multi-task link prediction task, we used a standard GraphSAGE layer and one cascaded MLP layer. 
We used Adam as the optimizer and $10^{-3}$ as the learning rate and trained all aspect forecasting and link prediction tasks for 200 epochs on a 16GB Nvidia-V100 GPU. 

\vspace{-3mm}
\subsection{Experimental results}

\subsubsection{Improvement on aspect forecasting without annotated aspects (RQ1)}
We first investigated the performance of ForeSeer on aspect forecasting when aspects are not annotated in the review (Table \ref{tab:main}). We found that our method achieved the best performance on all three pieces of time gaps (3 months, 6 months, and 3 years), indicating that our temporal graph embedding framework can effectively model the dynamics of aspects and products. For instance, ForeSeer obtains 63.1\% and 49.1\% improvements on 90-day gap forecasting and 3-year gap forecasting, respectively. We found that the frequency baseline didn't perform well on this task, especially when the time gap is larger, further confirming the importance of aligning products temporally. We found that the GNN baseline has undesirable performance, demonstrating the effectiveness of progressive training on avoiding over-fitting. 
We found that the MLP, GRU, and LSTM baselines performed badly on all future forecasting, necessitating the importance of graphical information. We also found that two temporal baselines outperform the MLP baseline, suggesting the importance to adjust product embedding for multi-time forecasting. We found that OA-Mine, one of the state-of-the-art approaches in aspect mining, obtained a less prominent performance, indicating that aspect mining methods cannot be applied to this novel aspect forecasting task. 
Finally, we noticed that graph-based approaches GNN, in general, performed better than methods that do not consider graphs, necessitating the consideration of graph dynamics in this task. 
\vspace{-2.1mm}

\subsubsection{Improvement on aspect forecasting with annotated aspects (RQ1, RQ2)}
We next evaluated an easier setting where aspects are annotated in each review. In this setting, the frequency baseline achieved very good performances since it simply counts the aspects in current reviews and uses it to predict future aspects. For aspects that are not time-sensitive, the frequency baseline can be regarded as an upper bound. Nevertheless, we found that ForeSeer still achieved good performance in this setting, where ForeSeer showed comparable performance with the frequency baseline on a 6-month gap and even outperformed the frequency baseline on a 3-year gap. All other comparison approaches obtained a much less promising performance. This indicates that ForeSeer can obtain a comparable performance with the frequency baseline on aspects that are not time-sensitive and substantially better performance on aspects that are time-sensitive. To further examine this (RQ2), we illustrated one case study showing how ForeSeer can effectively recognize long-term aspects (Table \ref{tab:case_study_1}). ForeSeer corrected forecasted time-sensitive aspects such as "a little goes a long way" and "sensitive skin", while the frequency baseline failed to identify, reassuring ForeSeer's ability to forecast long-term future aspects.  

\vspace{-4mm}
\begin{table*}[h]
    \centering
      \setlength{\belowcaptionskip}{-1mm}
    \caption{Comparison of ForeSeer and GNN on predicting future links at three different time points (current, 6 months, 3 years). }

    \scalebox{0.78}{
    \begin{tabular}{l|l|cccccc}
    \hline
         $\Delta t-$future  & Approaches & Accuracy &Precision & Recall& F1 score & AUPRC & AUROC \\
        \hline
        Current& Baseline GNN & 88.32± 0.54&   85.29 ± 0.37& 91.28 ± 0.72& 88.13 ± 0.61 & 94.64 ± 0.88 & 94.73 ± 0.76\\ 
          (0 day)&  \textbf{ForeSeer (Ours)}& \textbf{92.64 ± 0.53} & \textbf{92.68 ± 0.35}& \textbf{92.66 ± 0.69}& \textbf{92.65 ± 0.58} & \textbf{97.54 ± 0.83} &\textbf{97.66 ± 0.75}\\
         \hline
        \multirow{2}{*}{6 months} & Baseline GNN & 18.36 ± 0.32 & 10.93 ± 0.24& 89.02 ± 0.72& 19.49 ± 0.23 &89.13 ± 0.84 & 44.79 ± 0.24 \\
         &  \textbf{ForeSeer (Ours)}& \textbf{85.12 ± 0.51} &\textbf{92.96 ± 0.38} & \textbf{91.87 ± 0.78} &\textbf{ 92.37 ± 0.55} & \textbf{92.21 ± 0.87 }&\textbf{74.91 ± 0.49}\\
        \hline
        \multirow{2}{*}{3 years} 
         &  Baseline GNN & 80.28 ± 0.48& 77.95 ± 0.35 & 87 ± 0.63 & 82.03 ± 0.77 &92.24 ± 0.88 & 89.34 ± 0.81  \\
         &  \textbf{ForeSeer (Ours) }& \textbf{84.25 ± 0.51} & \textbf{92.95 ± 0.44} & \textbf{87.31 ± 0.88 }& \textbf{86.39 ± 0.64} & \textbf{94.08 ± 0.96}& \textbf{91.75 ± 1.02}\\
        \hline
    \end{tabular}
    }
    \label{tab:link_pred}
    \vspace{-4mm}
\end{table*}
\vspace{-0.5mm}

\subsubsection{Improvement on predicting review-aspect association with embedding matching (RQ3)}

\begin{figure}[h]
\centering
 \includegraphics[width=0.78\linewidth]{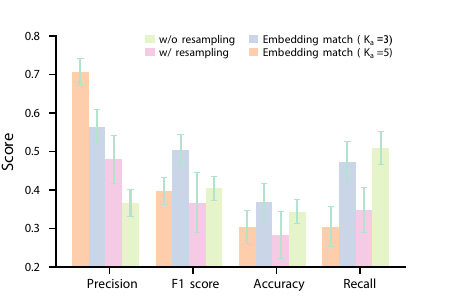}
\caption{Bar plot showing the performance of our embedding match approach and comparison approaches on review-aspect association prediction evaluated using precision, F1 score, accuracy, and recall.}
\label{fig:bert_barplot}
\vspace{-7mm}
\end{figure}

Next, we examined the importance of the aspect-guided embedding matching strategy by performing an ablation study on the review-aspect association prediction (Fig. \ref{fig:bert_barplot}). We observed the limited precision performance of using either with or without resampling multi-label classification instances, leading to poor ability or focusing too much on capturing tail aspect associations. Instead, we observed substantial improvements from our aspect-guided embedding framework by successfully exploiting the advantage of two multi-label classification instances. For example, the precision of our precision-inclined embedding matching model ($K_a = 5$) is 22\% and 34\% higher than the with or without resampling multi-label classification instances.
The superior precision performance suggests the learned review embeddings guided by aspects concentrate less-biased features for both head and tail aspects, which is critical to their success. 
\vspace{-5mm}

\subsubsection{Improvement on predicting future links (RQ4)}


After confirming the superior performance of ForeSeer on aspect forecasting, we next investigated whether ForeSeer's PATA module is able to capture temporal information for multiple futures by letting it predict product-product edges that might emerge in the multiple future times (Table \ref{tab:link_pred}). We observed a consistent improvement of our methods over different time gaps compared to GNN. We adopted a multi-task prediction setting for efficient inference and training where predictions for different time gaps are treated as different tasks. Specifically, we found that the GNN baseline cannot generalize well on the intermediate-long future range (6 months) as it didn't consider the dynamics when simultaneously predicting multiple time gaps. In contrast, our model addressed this by explicitly modeling the temporal dynamics of products, confirming that ForeSeer can effectively capture both product and aspect dynamics.

\vspace{-2mm}

\subsubsection{Contrasting static and temporal embeddings (RQ5)}
We further visualized the dynamic product embedding space colored by their product types at different time steps to assess how ForeSeer captures the temporal pattern (Fig. \ref{fig:UMAP_product}). As the time step becomes larger, we observed more visible clustering patterns among products. This suggests that the quality of the product embeddings becomes better as the data is accumulated by modeling the temporal dynamics. We also noticed the product embedding similarity reflects edges in the product-product graphs ("shelf" and "caddy", "lamp", "light fixture" and "string light"), indicating these product embeddings successfully encode graphical information.


%% file: main_paper/related_works.tex
\vspace{-3mm}
\section{Related works}
\noindent \textbf{Product aspect mining.} Few work that tries to forecast aspects for products and most previous works only aim at product aspect mining tasks. Mining product aspects from large-scale commercial data has been a well-studied problem \cite{hu2004mining,sharma2014mining, dadhich2021social, kpiebaareh2022generic, yang2022mave}. Previous works target the problem using rule-based \cite{sharma2014mining, dadhich2021social, kpiebaareh2022generic}, supervised learning based extractor \cite{jeyapriya2015extracting, chakraborty2022aspect} or propagation-based methods \cite{qiu2011opinion, lu2022aspect}. These works are less generalizable to product aspect forecasting, as they require either domain-specific features or supervision signals from downstream tasks such as sentiment or opinion labels.
Recently, other works similarly extract descriptive product attributes from seller-provided product profiles by using sequential supervised textual span labeling information \cite{xu2019scaling, zhu2020multimodal, wang2020learning, li2020emova, yan2021adatag} or distant supervision \cite{zhang2021discovering, wang2022attribute}. OA-Mine \cite{zhang2022oa} leverage weakly-supervised seed set to mine aspects from product titles. These methods need either more costly massive sequential labeled datasets or predesigned hierarchical attribute taxonomy, which is hard to obtain for customer-centered aspects. 
Zero-shot aspect extraction is another recently emerged related topic that aims to extract aspects in new domains without annotated data by leveraging transfer learning \cite{jebbara2019zero}, natural language inference \cite{shu2022zero}, document sentiment classification \cite{deng2022zero} and the recent emerged super-scale ChatGPT \cite{ouyang2022training} and GPT-4 \cite{bubeck2023sparks}. However, the zero-shot indicates that no prior knowledge of the new domain is needed instead of the number of reviews being limited for products. They are thus not directly applicable to forecasting tasks. In summary, all the methods above focus on text sequence level extraction and do not efficiently aggregate learned attributes at the product level. They are thus useful to help construct the aspect list for product aspect forecasting tasks, but are not able to forecast future aspect trends. Unlike these methods, our methods are the first framework that can forecast aspects that might be mentioned in future reviews for a new product with a limited number of reviews.

\noindent \textbf{Label-guided classification.} Label-guided text classification has been studied in fields such as social recommendation and document classification \cite{liu2017deep, chien2021node}. It has been proven to be beneficial for classification performance \cite{xiao2019label,ma2021label}. While these methods focus on classification, our method provides an efficient cross-domain contrastive learning framework that can easily ensemble these models as expert multi-label classification instances. Compared to \cite{wang2023multi}, which also proposes to mix multiple encoder instances, our representation learning scheme allows the model to also produce informative reviews and aspect representations to help downstream tasks.

\noindent \textbf{Temporal graph learning.} Temporal graph-based representation learning has been intensively studied \cite{rossi2020temporal, xue2022dynamic, guo2022continuous}. Previous works either incline on graph structure \cite{li2017attributed, zhu2018high, nguyen2018continuous, zhao2019large, schaefer2022aegnn} or temporal dynamics \cite{trivedi2017know, trivedi2019dyrep, ma2020streaming, min2021stgsn, xu2022dyng2g, jin2022stgnn}. The dynamic graph structure is extracted by graph adjacency dynamics \cite{li2018deep, goyal2020dyngraph2vec}, skip-gram-based modeling \cite{du2018dynamic, zuo2018embedding, mahdavi2018dynnode2vec}, leveraging clique information \cite{yu2018netwalk, zhou2018dynamic, huang2020motif}, step-by-step embedding updating \cite{zhu2016scalable, rahman2018dylink2vec, goyal2018dyngem, singer2019node, liu2019real, cai2021structural, liu2022dynamic}. These methods only focus on capturing implicit or short graphical dynamics changing and can not efficiently model large-scale review-aspect associations and evolving product-review associations. Some methods are designed specifically for heterogeneous networks \cite{zhang2017learning, wang2020dynamic, zhang2021tigecmn, ji2021temporal,zhan2022coarsas2hvec}.
Temporal-dynamic focused methods leverage recurrent-based models \cite{dai2016deep,zhu2017next, wu2017recurrent, beutel2018latent, kazemi2019time2vec}, continuous point process modeling \cite{sajadmanesh2019continuous, chang2020continuous} and self-supervised graph representation learning \cite{tian2021self,sun2022self}. These methods focus on modeling interactive events between nodes, but are not suitable for modeling dynamic product similarities and massive product-review associations. Some methods leverage self-attention, hierarchical, or temporal attention \cite{sankar2018dynamic, zheng2019addgraph, lu2019temporal, yang2020dynamic, xu2020inductive, xue2021modeling, jiao2021temporal}, but only for temporal information aggregation, while our PATA module leverages attention to predict multiple time range future simultaneously. Other recent works such as EvolveGCN \cite{pareja2020evolvegcn} and Roland \cite{you2022roland} incorporate recurrent modules and efficient graph learning. These frameworks can be easily adapted to our progressive training framework, while we aim to clearly show that progressive training is the key success to in aspect forecasting with a simple yet clear formulation. JODIE \cite{kumar2019predicting} is the closest work that can also predict multiple time-range futures with a time projection module. However, it is based on interaction events and is designed for heterogeneous user-item networks, making it impractical for our use case. In summary, most of the existing works can not be directly applied to the product aspect forecasting task because of the training efficiency and inability to handle large-scale reviews.

%% file: main_paper/conclusion.tex
\section{Conclusion}
In this paper, we have studied a novel task of product aspect forecasting, which aims to predict aspects that users might mention in future reviews. We have proposed a novel framework ForeSeer to solve this task by dynamically embedding products, reviews, and aspects. We have evaluated our method on a large-scale real-world dataset and observed ForeSeer's superior performance on aspect forecasting and link prediction. In the future, we are interested in boosting ForeSeer with more advanced progressive training strategies. We are interested in applying ForeSeer to other temporal graph embedding frameworks, such as modeling biological signaling pathways. We are further interested in exploring how ForeSeer can assist the classic task of aspect mining, allowing us to apply ForeSeer to advance a larger number of e-commerce applications, such as integrating it with recommendation systems to enhance the product suggestions based on anticipated aspect preferences.